\documentclass[a4paper,12pt,oneside,reqno]{amsart}
\usepackage{hyperref}
\usepackage[headinclude,DIV13]{typearea}
\areaset{15.1cm}{25.0cm}
\usepackage{txfonts,amssymb,amsmath,amsthm,bbm,dsfont}

\usepackage[utf8]{inputenc}
\usepackage{graphicx, psfrag}
\usepackage{xcolor}
\usepackage{subfigure}

\definecolor{bbm}{RGB}{51,153,0}
\definecolor{above}{RGB}{128,0,128}
\definecolor{below}{RGB}{102,0,204}
\definecolor{cascade}{RGB}{204,0,0}
\definecolor{iid}{RGB}{153,51,0}
\newcommand{\beginsupplement}{%
	\setcounter{section}{0}
        \setcounter{table}{0}
        \renewcommand{\thetable}{S\arabic{table}}%
        \setcounter{figure}{0}
        \renewcommand{\thefigure}{S\arabic{figure}}%
     }

\theoremstyle{remark}

\def\paragraph#1{\noindent \textbf{#1}}

\numberwithin{equation}{section}

\def\<{\langle}
\def\>{\rangle}

 \def \ba {\begin{array}}
 \def \ea {\end{array}}

 \newcommand{\be}{\begin{equation}}
 \newcommand{\ee}{\end{equation}}

\newcommand{\bea}{\begin{eqnarray}}
 \newcommand{\eea}{\end{eqnarray}}
\def\TH(#1){\label{#1}}\def\thv(#1){\ref{#1}}
\def\Eq(#1){\label{#1}}\def\eqv(#1){(\ref{#1})}

 \def \1{\mathbbm{1}}

\begin{document}

  %%%%%%%%% TITLE PAGE %%%%%%%%%%%%%%%%%%%%%%%%%%

 \title[Drop of prevalence after population expansion]{A lower prevalence for recessive disorders in a random mating population is a transient phenomenon during and after a growth phase}
 
 \author[L. La Rocca]{Luis A. La Rocca}
 \address{
 	L. La Rocca\\
	Institut f\"ur Angewandte Mathematik\\
	Rheinische Friedrich-Wilhelms-Universität\\
	Endenicher Allee 60\\
	53115 Bonn, Germany
	}
\email{luis.larocca@uni-bonn.de}

\author[J. Frank]{Julia Frank}
 \address{
 	J. Frank\\
	Institute for Genomic Statistics and Bioinformatics\\
	Rheinische Friedrich-Wilhelms-Universität\\
	Venusberg-Campus 1\\ 
	53127 Bonn, Germany }
\email{hjfrank@uni-bonn.de}

\author[H. B. Bentzen]{Heidi Beate Bentzen}
 \address{
 	H. B. Bentzen\\
	Faculty of Law, Norwegian Research Center for Computers and Law\\
	Oslo Univeristy \\
	Karl Johans gate 47 \\
	Domus Media \\
	0162 Oslo, Norway
	}
\email{h.b.bentzen@medisin.uio.no}

\author[J.-T. Pantel]{Jean-Tori Pantel}
 \address{
 	J.T. Pantel\\
 	Charité University Medicine Berlin \\
	Charitépl. 1 \\
	10117 Berlin, Germany}
\email{jean-tori.pantel@charite.de}

\author[K. Gerischer]{Konrad Gerischer}
 \address{
 	K. Gerischer\\
	Institute for Genomic Statistics and Bioinformatics\\
	Rheinische Friedrich-Wilhelms-Universität\\
	Venusberg-Campus 1\\ 
	53127 Bonn, Germany }
\email{kgerisch@rheinahrcampus.de}

\author[A. Bovier]{Anton Bovier}
\address{
	A. Bovier\\
	Institut f\"ur Angewandte Mathematik\\
	Rheinische Friedrich-Wilhelms-Universität\\
	Endenicher Allee 60\\
	53115 Bonn, Germany }
\email{bovier@uni-bonn.de}

\author[P. M. Krawitz]{Peter M. Krawitz}
 \address{
 	P. M. Krawitz\\
	Institute for Genomic Statistics and Bioinformatics\\
	Rheinische Friedrich-Wilhelms-Universität\\
	Venusberg-Campus 1\\ 
	53127 Bonn, Germany }
\email{pkrawitz@uni-bonn.de}

\date{\today}

 \begin{abstract}  
Despite increasing data from population-wide sequencing studies, the risk for recessive disorders in consanguineous partnerships is still heavily debated. An important aspect that has not sufficiently been investigated theoretically, is the influence of inbreeding on mutation load and incidence rates when the population sizes change. We therefore developed a model to study these dynamics for a wide range of growth and mating conditions. In the phase of population expansion and shortly afterwards, our simulations show that there is a drop of diseased individuals at the expense of an increasing mutation load for random mating, while both parameters remain almost constant in highly consanguineous partnerships. This explains the empirical observation in present times that a high degree of consanguinity is associated with an increased risk of autosomal recessive disorders. However, it also states that the higher frequency of severe recessive disorders with developmental delay in inbred populations is a transient phenomenon before a mutation-selection balance is reached again.
 \end{abstract}

\thanks{
This work was partially supported by the Deutsche Forschungsgemeinschaft (DFG, German Research
Foundation) under Germany’s Excellence Strategy GZ 2047/1, Projekt-ID 390685813 and GZ 2151,
Project-ID 390873048 and through the Priority Programme 1590 “Probabilistic Structures in Evolution”. We thank Benjamin C. Haller for his very helpful advices on how to get the best out of the forward genetic simulation framework SLiM \cite{Haller_Messer_2019}. 
 }

%\subjclass[xxx]{??, ??}

\maketitle

 %%%%%%%%% END TITLE PAGE %%%%%%%%%%%%%%%%%%%%%%%%%%

\section{Introduction and Main Result}

The empirical observation that consanguinity is associated with an increased risk of autosomal recessive disorders, has been made in many countries.
Martin, et al. recently showed that the contribution of autosomal recessive developmental disorders is 31\% in the current British population if the autozygosity is above $0.02$. \cite{Martin_Jones_McIntyre_2018}
Likewise, in the Iranian population it is estimated that offspring from first-cousin unions have a probability for intellectual disabilities that is four times higher than in non-consanguineous partnerships.
\cite{Hu_Kahrizi_2018,Kahrizi_Hu_Hosseini_2019,Musante_Ropers_2014}
Although most people probably agree that a lower burden of disease and child mortality is a desirable goal in a society, it is also clear that the occurrence and coexistence of different marriage patterns over many centuries cannot be understood by population genetics alone, especially as demographic, social and economic factors interact in a complex manner.
\cite{Bittles_Black_2010}
However, since there have been repeated attempts to motivate social conventions by genetic reasoning, we took a closer look at the validity of these arguments. 

The European Court of Human Rights case of St\"ubing v. Germany concerned consanguine siblings who had four children following consensual intercourse, whereupon both siblings were charged with incest.
\cite{Stuebing_v_Germany_2012}
One of the siblings lodged a complaint, arguing that the legislature violated his right to sexual self-determination, his private and family life.
The Court found that 24 out of 44 European States reviewed, criminalized consensual sexual acts between adult siblings, and all prohibited siblings from getting married.
The German government argued that the law against incest partly aimed to protect against the significantly increased risk of genetic damage among children from an incestuous relationship.
Motivating a law on avoiding a higher probability of disease can be viewed as eugenic.
As the German Ethics Council opined after the judgment, no convincing argument can be derived from there being a risk of genetic damage, concurring with a statement from the German Society of Human Genetics that ``The argument that reproduction needs to be thwarted in couples whose children possess an elevated risk for recessively inherited illnesses is an attack on the reproductive freedom of all''.
\cite{Deutscher_Ethikrat_2016, Eugenische_Argumentation_2008}
Furthermore, as our work shows, the argument that there exists an increased risk of genetic damage, requires the definition of a reference population for comparison.
However, there is neither agreement about a suitable reference nor an accurate measurement for mutation load.

Previously, the dynamics of mutation load and incidence rates have been analyzed for different demographic models including explosive population growth, but excluding different mating schemes.
\cite{Henn_Botigue_Bustamante_Clark_Gravel_2015}
In this work, we will therefore focus on simulations for random and consanguineous mating.
In this framework, the mutation load is defined as the average number of lethal equivalents per individual.
The lethal equivalents in the genome are deleterious alleles that are disease-causing if both copies of a gene in an individual harbor at least one such variant.
The totality of these mutations could also be regarded as the theoretical superset of an extended carrier screen.
\cite{Antonarakis_2019}
By this means, we are able to focus on the incidence rate of severe recessive disorders with early onset that prevent reproduction almost with certainty.
Likewise, we can study how the selection of a partner, which we refer to as a mating scheme, influences the disease prevalence and mutation load and we are able to monitor these parameters in the population over time.
This is done by counting the number of lethal equivalents that enter the gene pool due to a constant de novo mutation rate, or leave the gene pool due to selection.
If the disease prevalence does not change any more, the population is in a steady state, that is a flux balance for lethal equivalents.

In our simulations, affected individuals do not have a different life span from unaffected individuals, therefore prevalence and incidence are equivalent and their rate is proportional to the amount of lethal equivalents removed from the gene pool per generation.
In fact, the expected number of lethal equivalents that is lost by an affected individual that is not propagating is two.
This is equivalent to the difference in the average mutation load between affected and unaffected individuals and can also be derived from the simulations. 
An expansion of the population will affect prevalence and mutation load as we will discuss in more detail in the following.
However, the prevalence has to approach the same level for both mating schemes in the long run, as it is only constrained by parameters of genome biology.
In contrast, the mutation load is affected by the final population size.
Consanguineous mating goes along with a lower effective population size and therefore with a reduced capacity for lethal equivalents.
\cite{Kimura_Maruyama_Crow_1963}

To investigate the dynamics of mutation load and incidence rates under different mating schemes in growing populations we developed two models and ran independent simulations in different frameworks.
Each model had the advantage to handle certain aspects of nature particularly well.
The first is an adaptation of the classical Wright-Fisher model with discrete, non-overlapping generations run in the forward genetic simulation framework SLiM.
\cite{Fisher_1919,Haller_Messer_2019,Wright_1931}
The second is a diploid individual-based population dynamics model adapted from Bovier, et al.
\cite{Bovier_Coquille_Neukirch_2018}
In contrast to the Wright-Fisher model, individuals die and give birth at independent time points that are exponentially distributed.
Both models resulted in the same qualitative and comparable quantitative results which will be described in the following.
However, a detailed explanation of both models, the adaption of the ``discrete'' Wright-Fisher model, and the ``continuous'' model with adaptive dynamics can be found in the Supplemental Material.

\begin{figure}[ph]
	\label{fig:1}
	\centering
	\includegraphics[width=\textwidth]{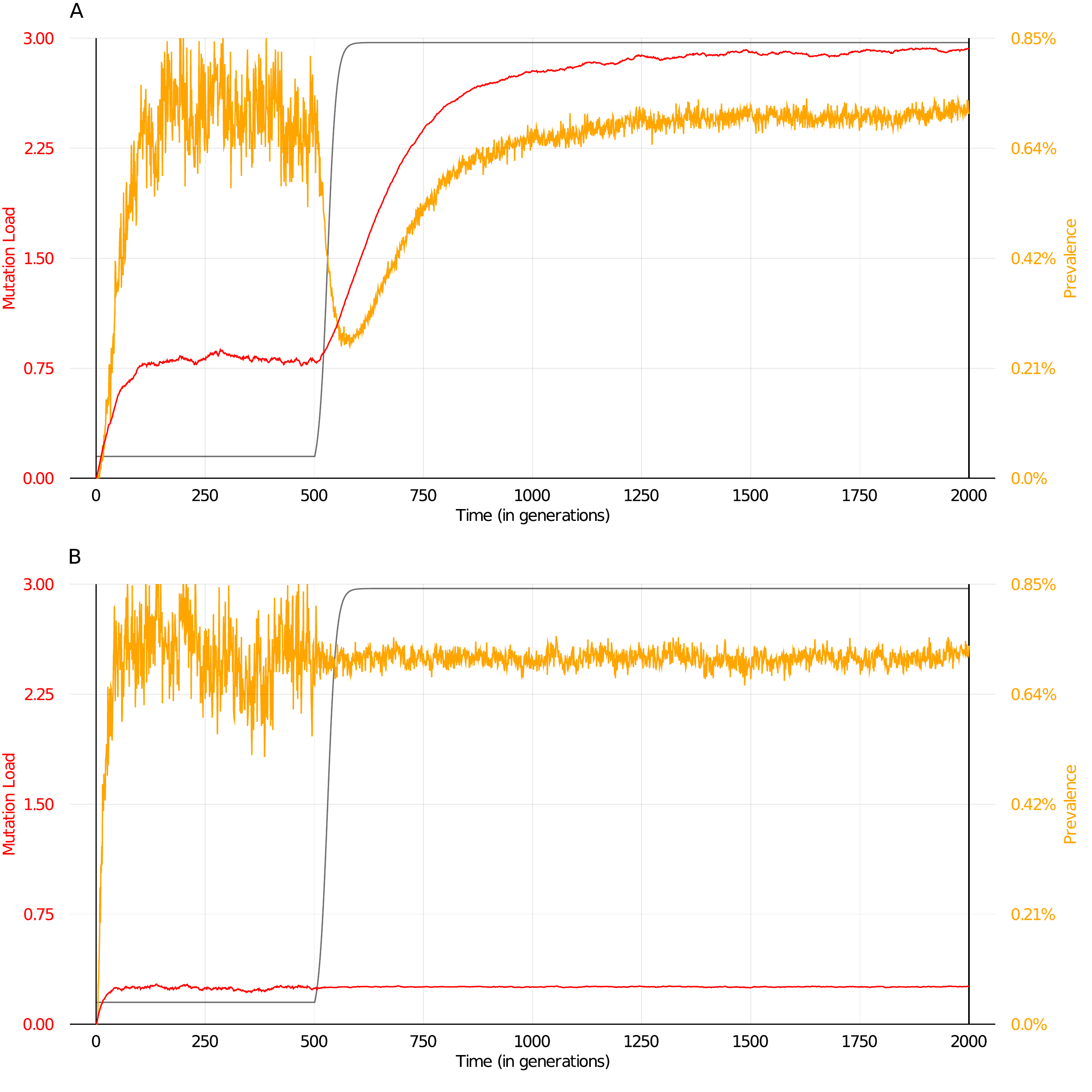}
	\caption{\textsc{Dynamics of mutation load and prevalence for severe recessive disorders}: A population expansion from 500 to $10\,000$ individuals (grey), starting in generation 500 does not affect prevalence (orange) nor mutation load (red) if partners are preferentially chosen within relatives (consanguineous mating scheme) (A). In contrast, in a random mating population, there is a transient drop of prevalence at the expense of an increasing mutation load (B). It takes more than 550 generations after the end of the growth phase, until the steady state is reached and the prevalence for both mating schemes are comparable again. The plots show the average of 50 exact trajectories of the stochastic process simulated with the discrete model.
}
\end{figure}

\begin{figure}[ph]
	\label{fig:2}
	\centering
	\includegraphics[width=\textwidth]{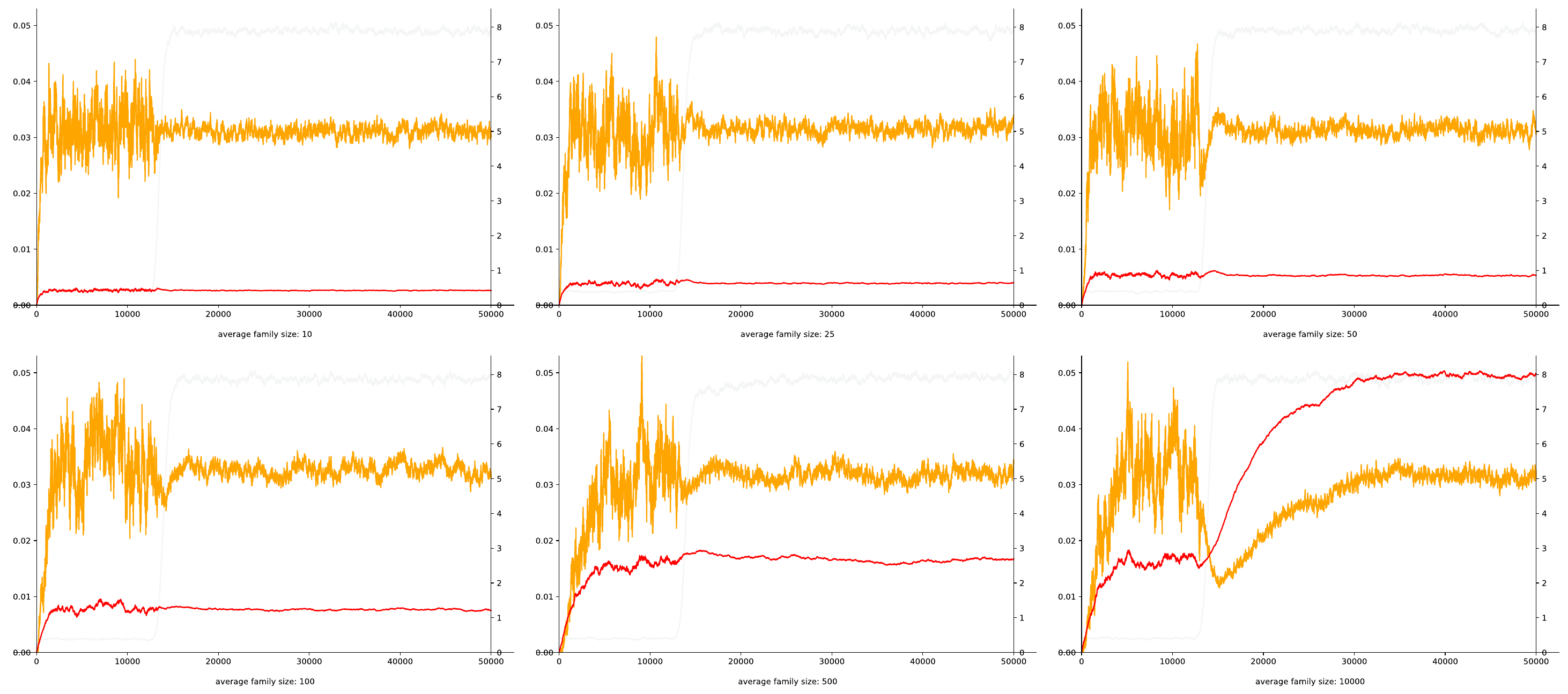}
	\caption{\textsc{Influence of family size on mutation load and prevalence}: The mating scheme is characterized by the family size and a probability function that describes how many of the partners are chosen within the family. In a preferentially consanguineous mating population the dynamics change when the maximum family size increases (upper left panel to lower right from 10, 25, 50, 100, 500 up to $10\,000$). The mutation load starts to increase considerably if mating is happening in tribes of 500 individuals. However, at this stage there is still only a minor effect of further population growth. In the lower right the maximum of the allowed family size is equivalent to the population size and thus, dynamics do not differ from a random mating scheme any more. The plots show the average of 10 exact trajectories of the stochastic process simulated with the continuous model.
}
\end{figure}

\begin{figure}[ph]
	\label{fig:3}
	\centering
	\includegraphics[width=\textwidth]{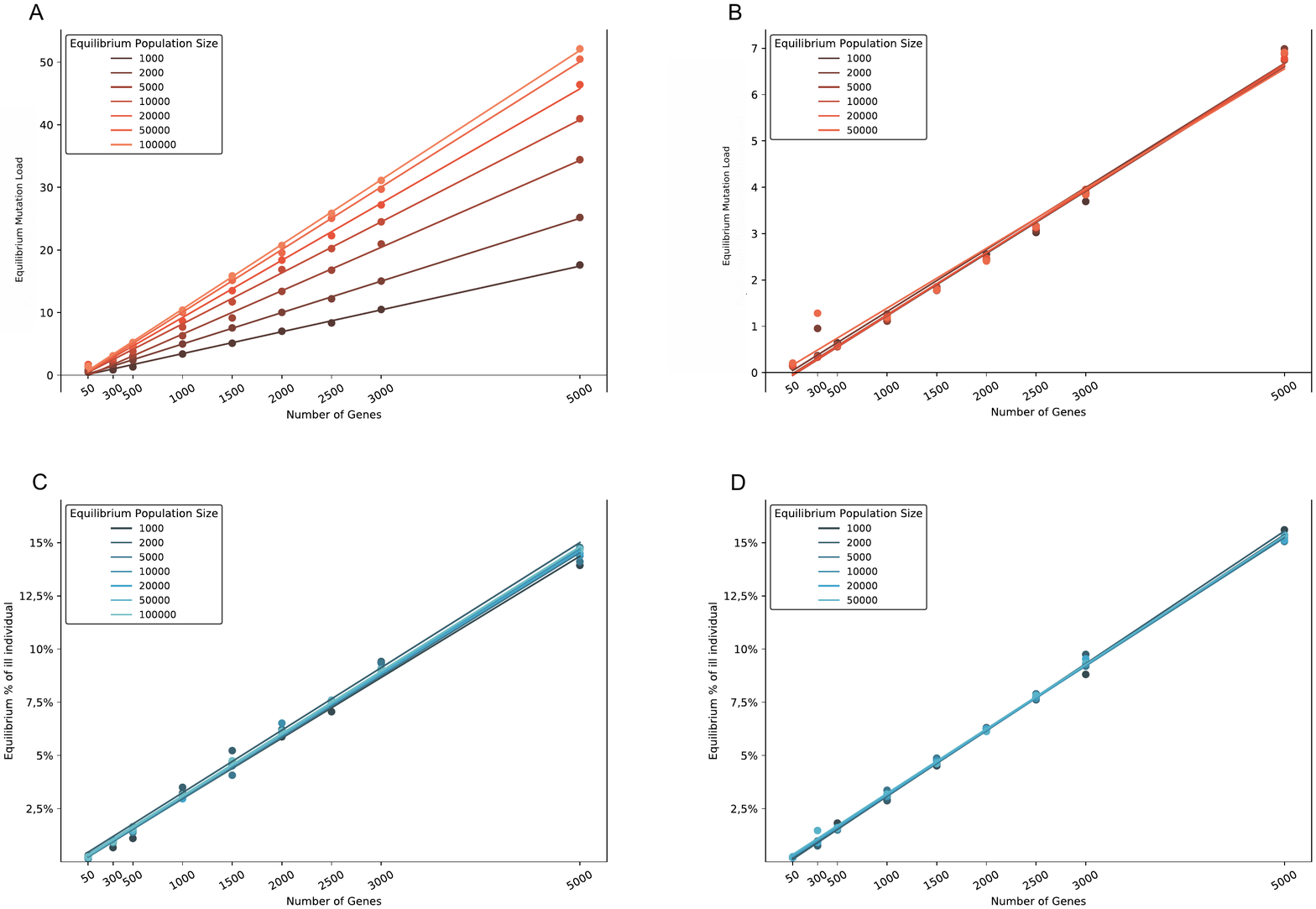}
	\caption{\textsc{Influence of genomic architecture and population size}: The capacity of the genome for deleterious mutations is larger in the random mating population. With an increasing number of genes and growing population size, deleterious mutations accumulate (A). In contrast, in the consanguineous mating scheme, family size limits the effective population size, and therefore mutation load is independent of the total number of individuals (B). Prevalence increases linearly in both mating schemes when the number of genes increases and is independent from population size, as regression analysis indicates (C,D).}
\end{figure}

We start our simulations with a population of 500 individuals that have around 500 generations to reach a steady state.
After this initialization phase, the population expands to a size of $10\,000$ in about 130 generations which would correspond to roughly $2\,500$ years.
\cite{Haller_Messer_2019}
The population expansion follows a logistic growth curve, which rather looks like a step function (grey curve in Figure \ref{fig:1}), because of the length of our entire simulations covering $2\,000$ generations.
All individuals have diploid genomes with $1\,000$ recessive genes that we deem crucial for reproductive success. Their coding sequence ranges between 500 and $10\,000$ bp, novel alleles are introduced with a de novo mutation rate of $1.2\cdot10^{-8}$ per bp, and one out of nine mutations is expected to be a lethal equivalent. The choice of these parameters can be motivated by the distribution of coding lengths and the deleteriousness scores for known autosomal recessive genes.
\cite{Kircher_Witten_Jain_ORoak_Cooper_Shendure_2014,Kochinke_Zweier_Nijhof_2016}
Pairs for procreation are formed either randomly or based on their relatedness that is traced over the two most recent generations. 
In a highly consanguineous mating scheme, the number of potential partners is hardly affected by the population size, as most marriages happen within families.
In our simulations this mating scheme is realized as follows:
$50\%$ of all partnerships share two grandparents, $30\%$ share one grandparent, and only $20\%$ share no grandparent.
In this scenario the mutation load and prevalence do not change during population growth (Figure \ref{fig:1} A).
In contrast, in a randomly mating population, there is a sharp transient drop of incidence rates during expansion at the expense of an increasing mutation load (Figure \ref{fig:1} B).
Interestingly, even after the final size of the population is reached, it takes almost another 550 generations until the mutation load reaches its new plateau of approximately three deleterious mutations in $1\,000$ recessive disease genes.
The prevalence, in contrast, returns to the value of the steady state, that is roughly 70 affected individuals per generation in a population of $10\,000$. 
The mutation load in the steady state increases in both mating schemes with the number of autosomal recessive genes, but with population size only for random mating (Figure \ref{fig:2} A, B).
This is best explained by a limit of the effective number of available partners that the consanguineous mating scheme imposes, regardless of the final population size.
In line with that argument is a transition from the dynamics of consanguineous to random when we incrementally increase family size, which would correspond to more potential mating partners (Figure \ref{fig:3}). 
Although the phase of population growth lasts only 130 generations in our simulations, the time span to reach the new equilibrium for the mutation load lasts much longer.

With respect to the British subpopulations of Pakistani (PABI) and European (EABI) ancestry in Martin, et al., this means that PABI are much closer to their steady state than EABI.
This would imply that the disease prevalence for recessive disorders will rather increase in the following generations for EABI, while it will remain constant on a high level for PABI. 
The subgroups of EABI and PABI that share a similar proportion of developmental delay due to de novo mutations, are best suited to discuss the mutational load: At first glance, the higher proportion of recessive coding mutations in the PABI subgroup with $>2\%$ autozygosity might indicate a higher mutational load.
However, our simulations would rather suggest the opposite.
Perhaps the higher mutation load in the EABI subgroup with multiple affected individuals is actually the key to understanding the high proportion of unexplained molecular causes in this subgroup.
\cite{Ji_Kember_Brown_Bucan_2016}
If oligogenic modes of inheritance explain a part of the undiagnosed patients, this share should be higher in the EABI subgroup. 

\newpage

\section{Figures}

\newpage

\newpage

\newpage

\section{Code Availability}
All scripts to reproduce our simulation results can be found in the following repository:

\begin{center} \url{https://github.com/roccminton/Diploid_Model_Two_Loci}. \end{center}

A video clip of our simulations can be found at:

\begin{center} \url{https://youtu.be/5hOgLyRqWPg}. \end{center}

\bibliographystyle{abbrv}
\bibliography{ms}

\newpage
\beginsupplement
\begin{center}
	\Large \textbf{\textsc{Supplemental Material}}
\end{center}
\vspace{1cm}

To investigate the dynamics of mutation load and incidence rates under different mating schemes in growing populations we developed two models and run independent simulations in different frameworks. 
Each has the advantage to model different aspects of nature better than the other.
One is an adaptation of the classical Wright-Fischer model with discrete, non overlapping generations \cite{Fisher_1919,Wright_1931} run in the forward genetic simulation framework SLiM \cite{Haller_Messer_2019}.
The other one is a diploid individual based population dynamics model adapted from Bovier et. al. \cite{Bovier_Coquille_Neukirch_2018}. 
In contrast to the Wright-Fischer model individuals die and give birth at independent exponential times on a continuous timescale. 
Therefore refer to the Wright-Fisher adaptation as ``discrete'' and to the adaptive dynamics model as ``continuous'' model.

\noindent Consider a finite population of individuals where each individual is characterized by a diploid set of \( N \) gene segments of different size. 
Mutations appear at every gene independently with a rate that is proportional to its size. 
As long as an individual carries a deleterious mutation at only one gene its fitness is unaffected. 
But as soon as both copies of a gene carry a mutation the individuals reproductive fitness is reduced to zero. 
Hence it will be excluded from the mating process and is not able to reproduce anymore.
One can think of a mutation causing a severe recessive disorder. 
Besides that all individuals are equally fit, no matter how many recessive disorders they cary.
Simulations always start with a small, healthy population. 
After a period of time in which a mutation selection balance is established a logistic growth phase starts, that settles after a new population equilibrium is reached. 
We are particularly interested in the evolution of two parameters during and after this growth phase. 
First the \textbf{mutation load}, which is the average number of mutations of an individual in the population. 
And second the \textbf{prevalence of the recessive disorder}, hence the fraction of the population that carries a mutation on both genes and thus expresses the disorder. 
Moreover we investigate changes of the dynamics of the above parameters when the population applies different mating schemes. On the one hand random mating, with an equal probability for each two individuals with non-zero fitness to mate. And on the other hand a consanguineous mating scheme, where individuals prefer to mate with close relatives.

\section{Discrete Model}
In the default setting, the simulation package from Haller and Messer \cite{Haller_Messer_2019}  samples a diploid population evolution according to the standard Wright Fisher model. 
Sexes were added such that each sex is equally represented in the population at any time.
In generation \( n \geq 1 \) there is a finite number of individuals \( M_n \geq 0 \) with a total of \( 2M_n \) genomes alive. 
In the initial burn-in phase the population size is held constant such that \( M_n = M_0 \) for all generations \( n \leq n_{grow} \). 
Afterwards the growth phase begins and the population size of each generation grows logistically with growth rate \( r > 0 \) until it approaches the carrying capacity \( K \). Therefore the population size of each generation is determined by the following formula
	\begin{equation*}
		M_n = \left\lceil \frac{K}{1+C_0 e^{-rK(n-n_{grow})}} \right\rceil \quad \text{ for all } n \geq n_{grow},
	\end{equation*}
where \( C_0 = \frac{K-M_0}{M_0} \).

\noindent The two mating schemes - random and consanguineous mating - are introduced as following. 
To create generation \( (n+1) \) first choose \( M_{n+1} \) women from generation \( n \) independently at random with replacement among all women with non-zero fitness. 
For random mating each woman then selects a man uniformly at random from the pool of potential partners with truly positive fitness.
To implement the consanguineous mating scheme use the pedigree information SLiM provides for the last two generations backwards in time.
Thus for every individual we know who are its parents and its grandparents. 
A woman in the consanguineous population now chooses a mate with a weighted uniform distribution on the set of all potential partners. For some weights \( \alpha, \beta \in [0,1] \) with \( \alpha + \beta \leq 1 \) the individual chooses a man with 
	\begin{align*}
		\text{two common grandparents} & \text{ with probability } \alpha \\
		\text{one common grandparent} & \text{ with probability } \beta \\
		\text{no common grandparents} & \text{ with probability } 1 - (\alpha + \beta)
	\end{align*}
Notice that having two grandparents is similar to being cousins and having one grandparent in common relates to being half-cousins. 

\noindent From the whole human genome pick \( N \) gene segments with an independent, uniformly distributed number of base pairs \( w_1, \dotsc , w_N \sim \mathcal{U}_{[a,b]} \), where \( a,b > 0 \) is the minimum resp. maximum segment size.
Moreover the whole genome is divided into \( n_c \) chromosomes.  
At birth not only mutation alters the offsprings genetic information, but also recombination.
For each chromosome start an independent \emph{Poisson} Process with rate \( r_\text{rec} > 0 \) marking the recombination breakpoints.
Here \( r_\text{rec} \) is the overall recombination rate.

\section{Continuous Model}
The adaptive dynamics model is continuous in time, hence time is not measured in \( n \in \mathbb{N} \) discrete generations, but on the positive real axis \( t \in \mathbb{R}_+ \). 
No exact pedigree are available for this continuous model, therefore introduce a new diploid family trait, which indicates the ancestry of an individual. 
Hence every individual is characterized by two diploid traits.
The first refers to the family origin whereas the second gives insight in the genetical information of the individual.
Introduce \( \mathcal{F} \subset \mathbb{N} \) as the finite set of all possible family traits and \( \mathbf{f} = (f^1,f^2) \in \mathcal{F}^2 \) being the family trait of an individual in the current population. 
Moreover the diploid genetic information of an individual is a finite vector
	\begin{equation*}
		\mathbf{x} = (x_1^1, \dotsc , x_1^N, x_2^1, \dotsc , x_2^N) \in \mathbb{N}_{\geq 0}^{2N}
	\end{equation*}
where the entries \( x_1^ i \) and \( x_2 ^i \) represent the number of mutation at the \( i^\text{th} \) gene segment in the first and second genome.
Thus if there are \( (\mathbf{f}_1,\mathbf{x}_1), \dotsc , (\mathbf{f}_{M_t},\mathbf{x}_{M_t}) \) individuals alive at time \( t > 0 \) in an arbitrary order, define the population state as a point measure on \( \mathcal{X} \coloneqq \left( \mathcal{F}^2 \times \mathbb{N}^{2N} \right) \) 
	\begin{equation*}
		\nu_t(\cdot) \coloneqq \sum\limits_{i = 1}^{M_t} \delta_{(\mathbf{f}_i,\mathbf{x}_i)}\left( \cdot \right)
	\end{equation*}
Individuals give birth and die at exponential rates \( b(\mathbf{f}, \mathbf{x}) \) resp. \(  d(\mathbf{f}, \mathbf{x}) \) which depend on the family trait and the chromosomal configuration of the individuals.
One can think of every individual carrying two independent clocks, a birth and a death clock. 
If the birth clock rings first the individual makes its mating choice, reproduces and resets its clock. 
Whereas if the death clock rings the individual disappears from the population.
Additional to the intrinsic death rate every individual sense the competition pressure of every other individual in the population. The term \( C(\mathbf{f}, \mathbf{x},\mathbf{g}, \mathbf{y}) \) gives the competition pressure executed by an individual of type \( (\mathbf{g}, \mathbf{y}) \) and felt by an individual of type \( (\mathbf{f}, \mathbf{x}) \). Hence the total death rate of an individual in population \( \nu \) is increased by the term 
	\begin{equation*}
		\int\limits_{\mathcal{X}} C(\mathbf{f},\mathbf{x},\mathbf{g},\mathbf{y}) d\nu(\mathbf{g},\mathbf{y})
	\end{equation*}
When an individual gives birth it chooses a partner at random from the population according to the partners birth rate and the reproductive compatibility between them. 
Notice that the reproductive compatibility \( R_{\mathbf{f}}\left(\mathbf{g}\right) \in [0,1] \)  of two individuals depends only on their family traits \( \mathbf{f} \) and \( \mathbf{g} \). 
After a mate was chosen the newborns family trait will be a uniform random combination of the four parental family traits unless both parents have the same traits. 
In this case the child inherits the exact same couple of traits. 
The genetic configuration of the newborn is not only a random combination of the parental alleles since the effect of mutation and the reshuffling of the parental chromosomes come into play. 
For an accurate definition of the reshuffling of the diploid parental chromosomes to a mixed haploid set, define first the sections on the genetic information, that form chromosomes.
Let \( n_c \in \mathbb{N} \) be the number of chromosomes for every individual.
Introduce the chromosome breakpoints  \( \{c_1, c_2, \dotsc ,c_{n_c + 1} \} \in \{1, \dotsc , N \} \) with \( 0 = c_1 < c_2 < \cdots < c_{n_c} < c_{n_c+1} = N \).
Divide the genetic information of every individual \( \mathbf{x} = (x_1^1, \dotsc , x_N^1, x_1^2, \dotsc , x_N^2) \) into chromosomes of the same length in both copies
	\begin{equation*}
		\mathbf{x} = (\underline{x}_1^1, \dotsc ,\underline{x}_{n_c}^1, \underline{x}_1^2 , \dotsc , \underline{x}_{n_c}^2) \quad \text{with} \quad \underline{x}_i^j = \left(x_{c_i + 1 }^j , x_{c_i+2}^j , \dotsc , x_{c_{i+1}}^j\right) 
	\end{equation*}
\noindent for \( j \in \{1,2\} \) and \( i \in \{1,\dotsc,n_c\} \). Finally get the reshuffled gamete from \( \mathbf{x} \) via a selection variable \( \tau : \{1,2, \dots , n_c \} \longrightarrow \{1,2\} \) as follows
	\begin{equation*}
		\mathbf{x}^\tau \coloneqq \left(\underline{x}_1^{\tau(1)} , \dotsc , \underline{x}_{n_c}^{\tau(n_c)} \right)
	\end{equation*}
Selecting \( \tau \) among all possible assignments equals an uniform recombination of the diploid chromosomes into haploid set.
At each birth a \emph{Poisson} distributed number of mutations is added to every offspring. 
The expectation of these identically distributed and independent \emph{Poisson} random variables equals the total mutation rate \( 2 \mu \bar{w} \), where \( \mu \) is the mutation rate per base pair and \( \bar{w} = w_1 + \dots + w_N \) is the sum of the length of the gene segments under consideration.
After sampling the number of mutations per birth, these mutations are distributed equally on the \( 2N \) gene segments according to their length.
Finally a mutation at a given gene increases the genetic value by one. 
All of this is captured in the mutation with recombination operator \( M^{\mathtt{rec}}_{\mathbf{f},\mathbf{x},\mathbf{g},\mathbf{y}} \) for a mating between individuals \( (\mathbf{f},\mathbf{x}) \) and \( (\mathbf{g},\mathbf{y}) \)
	\begin{equation*}
		\left( M^{\mathtt{rec}}_{\mathbf{f},\mathbf{x},\mathbf{g},\mathbf{y}}\phi \right) \! (\nu) \! = 
		\begin{cases}
			\dfrac{1}{2^{2n_c}} \! \sum\limits_{\tau, \tau^\prime \in \{1,2\}^{n_c}} \! \int\limits_{\mathcal{X}} \! \left( \! \phi\left( \nu \! + \! \delta_{\mathbf{f},\left(\mathbf{x}^\tau, \mathbf{y}^{\tau^\prime}\right) + h} \right) \! - \! \phi(\nu) \right) \! m\left( (\mathbf{x},\mathbf{y}) , dh \right) \text{ ,for } \mathbf{f} = \mathbf{g} \\ 
		\dfrac{1}{2^{2n_c+2}} \! \sum\limits_ {\substack{i,j \in \{1,2\} \\ \tau, \tau^\prime \in \{1,2\}^{n_c}}} \! \int\limits_{\mathcal{X}} \! \left( \! \phi\left( \nu \! + \! \delta_{f_i,g_j,\left(\mathbf{x}^\tau, \mathbf{y}^{\tau^\prime}\right) + h} \right) \! - \! \phi(\nu) \right) \! m\left( (\mathbf{x},\mathbf{y}) , dh \right) \text{ ,else} 
		\end{cases}
	\end{equation*}
and the mutation measure \( m \) is defined as
	\begin{equation*}
	m\left( (x,y) , dh \right) \coloneqq \sum\limits_{k=0}^\infty \left( \frac{(2\mu\bar{w})^k e^{-2\mu\bar{w}}}{k!} \frac{1}{Z_k} \sum\limits_{l \in \Diamond_k^{2N}} \left( \sum\limits_{i=1}^N \left( w_i^{l_i} \mathds{1}_{\{ l_i > 0 \}} + w_i^{l_{i+N}} \mathds{1}_{\{ l_{i + N} > 0 \}} \right) \right) \delta_l(dh) \right)
	\end{equation*}
\noindent where \( \Diamond_k^{2N} \coloneqq \left\{ l \in \mathbb{N}^{2N}_{+} \mid l_1 + \cdots + l_{2N} = k \right\} \) is the set of all lattice vectors in \( \mathbb{N}_{+}^{2N} \) with one norm equal to \( k \) moreover \( Z_k > 0 \) is a normalizing constant depending on the size of \( \Diamond_k^{2N} \). 
Let 
	\begin{equation*}
		\mathcal{M}(\mathcal{X}) = \left\{ \sum\limits_{i=1}^n \delta_{\mathbf{f}_i,\mathbf{x}_i} \colon n \geq 0, (\mathbf{f}_1,\mathbf{x}_1), \dotsc , (\mathbf{f}_n,\mathbf{x}_n) \in \mathcal{X} \right\}
	\end{equation*}
be the set of all finite point measures on \( \mathcal{X}\).
The dynamics of the continuous time, \( \mathcal{M}(\mathcal{X}) \)-valued jump process \( \left( \nu_{t} \right)_{t \geq 0} \) can be described by the generator \( \mathcal{L} \) defined for any bounded measurable function \( \phi \colon \mathcal{M}(\mathcal{X}) \longrightarrow \mathbb{R} \) as
	\begin{align*}
		\mathcal{L}\phi(\nu) = &\int\limits_{\mathcal{X}} b(\mathbf{f},\mathbf{x}) \left( \int_{\mathcal{X}} \frac{b(\mathbf{g},\mathbf{y})R_{\mathbf{f}}\left(\mathbf{g}\right)}{\langle \nu ,b \cdot R_{\mathbf{f}} \rangle} \left(M^{\mathtt{rec}}_{\mathbf{f},\mathbf{x},\mathbf{g},\mathbf{y}}\phi \right) (\nu) d\nu(\mathbf{g},\mathbf{y}) \right) d\nu(\mathbf{f},\mathbf{x}) \\
		& + \int\limits_{\mathcal{X}} \left( d(\mathbf{f},\mathbf{x}) + \int\limits_{\mathcal{X}} C(\mathbf{f},\mathbf{x},\mathbf{g},\mathbf{y}) d\nu(\mathbf{g},\mathbf{y}) \right) \left( \phi \left( \nu - \delta_{(\mathbf{f},\mathbf{x})}) - \phi(\nu) \right) \right) d\nu(\mathbf{f},\mathbf{x}) \\
	\end{align*}
Assume the boundedness of the birth and death rates \( b \) and \( d \) as well as the boundedness of the competition kernel \( C \). Starting in a initial state \( \nu_0 \in \mathcal{X} \) such that
	\begin{equation*}
		\mathbb{E}\left[ \nu_0, 1 \right] < \infty
	\end{equation*}
existence and uniqueness in law of the process with infinitesimal generator \( \mathcal{L} \) and initial condition \( \nu_0 \) can be derived from \cite{Fournier}.

\paragraph{Keep families small}
To ensure a stable mating scheme during the evolution of the population split families which became too big up into two subfamilies. Therefore introduce the following sequence of stopping times.
	\begin{align*}
		\theta_0 & \coloneqq 0  \\
		\theta_{k+1} & \coloneqq \inf\left\{ t>\theta_{k} : \exists \mathbf{f} \in \mathcal{F}^2 \text{ s.th. } \langle \nu , \mathds{1}_{\mathbf{f}} \rangle > \kappa \right\}
	\end{align*}
for some fixed \( \kappa > 0 \). Notice that at any stopping time it is always a unique family \( \mathbf{f} \in \mathcal{F}^2 \) that exceeds the maximum family size, since at any time there is at most one individual entering or exiting the population. At these random times the big family is split up at random into two subfamilies, where the size of each subfamily is binomial distributed with mean \( \frac{1}{2} \). One family keeps the old family trait and the other one gets a completely new, homogeneous one. To make this precise associate a number to each individual in a family. Therefore let \( H^{\mathbf{f}} = (H^{\mathbf{f}}_1, H^{\mathbf{f}}_2 , \dotsc , H^{\mathbf{f}}_k, \dotsc  ) : \mathcal{M}(\mathcal{X}) \longrightarrow (\mathbb{N}^{2N})^{\mathbb{N}} \) defined by 
	\begin{equation*}
		H^{\mathbf{f}} \left( \sum\limits_{i=1}^n \delta_{\mathbf{g}_i,\mathbf{x}_i} \right) = \left( \mathbf{x}_{\sigma(1)}, \mathbf{x}_{\sigma(2)}, \dotsc ,\mathbf{x}_{\sigma(k)}, 0, 0, 0, \dotsc  \right)
	\end{equation*}
where the \( \mathbf{x}_i \) for \( i=1,\dotsc,k \) are the genetic configurations of the individuals in \( ((\mathbf{g}_1,\mathbf{x}_1) \) \(, \dotsc ,\) \( (\mathbf{g}_n,\mathbf{x}_n)) \) with \( \mathbf{g}_i = \mathbf{f} \) and where \( \mathbf{x}_{\sigma(1)} \preceq \dots \preceq \mathbf{x}_{\sigma(k)} \) is the lexicographical order \( \preceq \) on \( \mathbb{N}^{2N} \) and \( k = \langle \nu , \mathds{1}_{\mathbf{f}} \rangle \) is the family size of \( \mathbf{f} \). Then the splitting of a family with trait \( \mathbf{f} \in \mathcal{F}^2 \) can be expressed with the following operator for any bounded measurable function \( \phi \colon \mathcal{M}_F(\mathcal{X}) \longrightarrow \mathbb{R} \)
	\begin{equation*}
		\left( S_\mathbf{f} \phi \right) (\nu) \coloneqq \frac{1}{2^{\langle \nu , \mathds{1}_\mathbf{f} \rangle}} \sum\limits_{\pi \in \{0,1\}^{\langle \nu , \mathds{1}_{\mathbf{f}} \rangle}} \left( \phi \left( \nu + \Delta_{\nu,\mathbf{f}}(\pi) \right) - \phi(\nu) \right) 
	\end{equation*}
where \(  \Delta_{\nu,\mathbf{f}}(\pi) \) executes the splitting of the family \( \mathbf{f} \) in \( \nu \) into two with configuration \( \pi \), hence
	\begin{equation*}
	 \Delta_{\nu, \mathbf{f}}(\pi) \coloneqq \sum\limits_{i=1}^{\langle \nu , \mathds{1}_{\mathbf{f}} \rangle} \pi(i) \left( - \delta_{\mathbf{f},H_i^{\mathbf{f}}} + \delta_{\mathbf{f}^{\mathtt{new}},H_i^{\mathbf{f}}} \right) d\nu(\mathbf{f},\mathbf{y})
\end{equation*}
where \( \mathbf{f}^{\mathtt{new}} = (f^{\mathtt{new}},f^{\mathtt{new}}) \) with \( f^{\mathtt{new}} \in  \{ g \in \mathcal{F} \mid \langle \nu , \mathds{1}_{\left(g,g^\prime\right)}\rangle = 0 , \forall g^\prime \in \mathcal{F}\} \) chosen deterministically, is a homogeneous family trait which is entirely new to the population. A possible way of choosing the new family trait at time \( \theta_k \) is to set  \( f^{\mathtt{new}} = n_{\mathcal{F}} + k \) where \( n_{\mathcal{F}} \) is the number of different families in the initial population \( \nu_0 \).
Hence the dynamics of the evolutionary process with splitting is given as the solution of the following martingale problem. Let \( \nu_0 \in \mathcal{M}(\mathcal{X}) \) be a initial population then for any real, continuous, bounded function \( \phi \) on \( \mathcal{M}(\mathcal{X}) \) the process
	\begin{equation*}
		M_t^{\phi} = \phi(\nu_t) - \phi(\nu_0) - \left( \int\limits_0^t \mathcal{L}\phi(\nu_s) ds + \sum\limits_{\mathbf{f} \in \mathcal{F}^2} \left( S_{\mathbf{f}} \phi \right)(\nu_t) \mathds{1}_{\langle \nu_t , \mathds{1}_{\mathbf{f}} \rangle > \kappa} \right)
	\end{equation*}
is a martingale.
\paragraph{Choices of Parameters}
Introduce the subset \( \mathcal{D}_N \subseteq \mathcal{X} \) of traits having at least one mutation in the same gene segment on both copies of the chromosome as
	 \begin{equation*}
		\mathcal{D}_N \coloneqq \left\{ (\mathbf{f},\mathbf{x}) \in \mathcal{X}^2 \, \bigg \vert \, \exists n \in \left\{ 1, \dotsc ,N \right\} \text{ s.th. } x^1_n > 0 \text{ and } x^2_n > 0 \right\}
	\end{equation*}
and set the birth and death rate to constant unless an individual falls in the set of non-propagable types 
	\begin{equation*}
		b(\mathbf{f},\mathbf{x}) = \bar{b} \mathds{1}_{(\mathbf{f},\mathbf{x}) \notin \mathcal{D}_L} \quad , \quad d(\mathbf{f},\mathbf{x}) = \bar{d}
	\end{equation*}
for \( \bar{b} > \bar{d} > 0 \).
To ensure uniform competition among all individuals set the competition pressure constant to
	\begin{equation*}
		 C(\mathbf{f}, \mathbf{x},\mathbf{g}, \mathbf{y}) = \frac{\bar{b}-\bar{d}}{K}
	\end{equation*}
for all individuals \( (\mathbf{f}, \mathbf{x}) \) and \((\mathbf{g}, \mathbf{y}) \). Therefore the population size in equilibrium fluctuates around the carrying capacity \( K \) of the system.
The different mating schemes are defined as follows. First the random mating, where \(R^{\mathtt{rnd}}_\mathbf{f}(\mathbf{g}) = 1 \) for all \( \mathbf{f}, \mathbf{g} \in \mathcal{F}^2 \) and the consanguineous mating scheme with
	\begin{equation*}
		R^{\mathtt{cng}}_{(f_1,f_2)}(g_1,g_2)  \coloneqq 
			\begin{cases} 
				\frac{2\alpha}{\kappa} & \text{if } (f_1,f_2) = (g_1,g_2) \\
				\frac{1-\alpha}{K-\frac{\kappa}{2}} & \text{else }
			\end{cases} 
	\end{equation*}
where \( \alpha \in [0,1) \) and \( \kappa > 0 \) is the maximum family size. Therefore the probability of mating within their own family that is of size \( \frac{\kappa}{2} \) in a population that is at its stable equilibrium size \( K \) is constant \( \alpha \). Note that for families with family size smaller than \( \frac{\kappa}{2} \) the probability of mating within the family is slightly lower, whereas it gets bitter when the family size surpasses this size. Furthermore the probability increases at the beginning of the growth phase when the carrying capacity \( K \) gets uplifted and the population size starts to grow slowly. During this initial phase of expansion there will be more consanguineous mating overall. This imbalance levels off as soon as the population size approaches \( K \).

\noindent Start the evolution with a small, healthy population of size \( M_0 > 0 \) in population equilibrium, where nobody carries any mutation.
The clonal individuals are following divided into \( n_{\mathcal{F}} \in \mathbb{N}_{>0} \) families with homogeneous family traits \( (1,1), (2,2), \dotsc ,(n_{\mathcal{F}},n_{\mathcal{F}}) \). 
After an initial phase during which a mutation selection balance is established raise the carrying capacity to generate a natural exponential population growth up to the new equilibrium.
The population parameters we are particular interested in, the mutation load and the prevalence rate for the disability can be formulated in terms of the population process.
The relative mutation load of a population \( \nu \) is defined as
	\begin{equation*}
		L(\nu) \coloneqq \frac{1}{\langle \nu , 1 \rangle} \int_{\mathcal{X}} l(\mathbf{x}) d\nu(\mathbf{f},\mathbf{x}) \quad \text{ with } \quad l(\mathbf{x}) \coloneqq \sum\limits_{i\in\{1,2\}} \sum\limits_{n=1}^N x_n^i
	\end{equation*}
And the relative number of individuals in the population \( \nu \) belonging to the set \( \mathcal{D}_N \) is
	\begin{equation*}
		I(\nu) \coloneqq \frac{\nu(\mathcal{D}_N)}{\langle \nu , 1 \rangle}  
	\end{equation*}
To generate stochastically correct trajectories of the population dynamics process we implemented a variation of Gillespie algorithm for the above model in Python.

\section{Comparing both models}
Both models, the discrete generation model implemented with SLiM and the continuous time model with the Gillespie algorithm, fit different aspects of nature better than the other.
A key feature of SLiM and the discrete model is the exact pedigree information generated for every individual. 
Whereas the continuous model can only cluster roughly in family clans, but cannot differ between members of one family.
A major drawback of the discrete model are the non overlapping generations.
That excludes the possibility of matings between individuals on different pedigree levels such as uncle - niece marriages.
This deficit is set aside by the continuous time model. Since individuals give birth and die independently, all individuals are of different age, and thus different generations are alive simultaneously.
On the one hand alike the Wright-Fisher model, the discrete model works with constant respectively deterministically increasing population sizes. 
On the other hand the continuous model has a fluctuating and natural growing population. 
Notice that for large populations the random fluctuation in population size are of order \( \frac{1}{K} \) and the stochastic process converges in law to the solution of a deterministic, logistic equation \cite{Fournier}.

\noindent Recombination is also implemented differently.
As SLiM works with true inter chromosomal recombination the continuous model only reshuffles the parental chromosomes while producing the gamete.
This is due to the fact that SLiM saves the exact base position of mutations on the human genome, whereas the continuous model oxnly knows the number of mutations per gene segment and not their exact location within each segment. 
Under the assumption that all genes are compound heterozygote the different implementation of recombination has not affect on the fitness of individuals.

\noindent Besides all differences parameters are chosen to be equal for both simulations. Among these the number of gene segments \( N \), the initial and equilibrium population size \( M_0 \) and \( K \), and many more.
Moreover family sizes in the continuous model are chosen such that the number of potential partners in both models in the consanguineous setting are approximately equal. 
Likewise the birth rate in the continuous time model is set to \( b = 1 \), such that in generation \( t \) there are \( M_{t+1} \) birth events, where \( M_t \) is the population size at that time. 
The only difference is, that for the discrete generation model there are \emph{exactly} \( M_{t+1} \) birth, whereas in the continuous time model there are \emph{on average} that many births

\end{document}